\documentclass [prl, twocolumn, superscriptaddress]{revtex4-1} 
\usepackage[dvipdfmx]{graphicx}
\usepackage{graphicx}
\usepackage{physics}
\usepackage{bm, amsmath, amssymb, braket}
\usepackage{times}
\usepackage{multirow}
\usepackage{ascmac}
\usepackage{amsthm}
\usepackage{float}
\usepackage{color}
\usepackage{comment}

\begin{document}

\title{Continuous Phase Transition without Gap Closing in Non-Hermitian Quantum Many-Body Systems}

\author{Norifumi Matsumoto}
\email{matsumoto@cat.phys.s.u-tokyo.ac.jp}
\affiliation{Department of Physics, University of Tokyo, 7-3-1 Hongo, Bunkyo-ku, Tokyo 113-0033, Japan}

\author{Kohei Kawabata}
\affiliation{Department of Physics, University of Tokyo, 7-3-1 Hongo, Bunkyo-ku, Tokyo 113-0033, Japan}

\author{Yuto Ashida}
\affiliation{Department of Physics, University of Tokyo, 7-3-1 Hongo, Bunkyo-ku, Tokyo 113-0033, Japan}
\affiliation{Institute for Physics of Intelligence, University of Tokyo, 7-3-1 Hongo, Bunkyo-ku, Tokyo 113-0033, Japan}

\author{Shunsuke Furukawa}
\affiliation{Department of Physics, Keio University, 3-14-1 Hiyoshi, Kohoku-ku, Yokohama, Kanagawa 223-8522, Japan}

\author{Masahito Ueda}
\affiliation{Department of Physics, University of Tokyo, 7-3-1 Hongo, Bunkyo-ku, Tokyo 113-0033, Japan}
\affiliation{Institute for Physics of Intelligence, University of Tokyo, 7-3-1 Hongo, Bunkyo-ku, Tokyo 113-0033, Japan}
\affiliation{RIKEN Center for Emergent Matter Science (CEMS), Wako 351-0198, Japan}

\date{\today}

\begin{abstract}

Contrary to the conventional wisdom in Hermitian systems, a continuous quantum phase transition between gapped phases is shown to occur without closing the energy gap $\Delta$ in non-Hermitian quantum many-body systems. 
Here, the relevant length scale $\xi \simeq v_{\rm LR}/\Delta$ diverges because of the breakdown of the Lieb-Robinson bound on the velocity (i.e., unboundedness of $v_{\rm LR}$) rather than vanishing of the energy gap $\Delta$. The susceptibility to a change in the system parameter  
exhibits a singularity due to nonorthogonality of eigenstates. As an illustrative example, we present an exactly solvable model by generalizing Kitaev's toric-code model to a non-Hermitian regime.
\end{abstract}

\maketitle

%%%%% Introduction %%%%%

Quantum phase transitions have long been a subject of active research in quantum many-body physics. A quantum phase is characterized by the low-energy and long-distance properties of a system such as the decay behavior of correlation functions of local operators in the ground state, the ground-state degeneracy, and its stability against local perturbations~\cite{Hastings2005}. At the transition
point
between different quantum phases, physical quantities show singularities reflecting changes in the long-distance behavior~\cite{Sachdev2001}. 
For conventional quantum many-body systems described by local and Hermitian Hamiltonians, it is widely accepted that a continuous quantum phase transition between gapped phases is accompanied by closing of an excitation gap $\Delta$. 
This correspondence is %one of 
among
the most fundamental properties of continuous phase transitions, and two gapped ground states %which 
that
are connected without gap closing are generally considered to belong to the same quantum phase~\cite{Chen2010}. This implies that long-distance properties of a ground state are preserved under continuous deformation of a local and gapped Hamiltonian. In fact, a change in the ground state under such %a 
deformation can be represented as a finite-time evolution generated by a local effective Hamiltonian, which preserves the long-distance structure of the ground state~\cite{Hastings2005, Bravyi2010, Chen2010}.

Meanwhile, non-Hermitian physics~\cite{Bender1998, Bender2002, Bender2007} has recently attracted widespread attention~\cite{Konotop2016, Feng2017, El-Ganainy2018, Miri2019, Ozdemir2019}. Non-Hermiticity originates from gain and loss of energy or particles in classical systems~\cite{Makris2008, Klaiman2008, Guo2009, Ruter2010, Lin2011, Regensburger2012, Peng2014}, and non-Hermitian quantum dynamics is realized under continuous observation without quantum jumps~\cite{Dalibard1992, Carmichael1993, Brody2012, Daley2014, Lee2014X, Lee2014L, Ashida2016, Ashida2017, Xiao2017, Ashida2018, Li2019, Wu2019, Yamamoto2019}. Some fundamental principles in Hermitian systems break down in non-Hermitian systems. Even in single-particle problems, unique topological phases~\cite{Rudner2009, Esaki2011, Schomerus2013, Leykam2017, Xu2017, Kozii2017, Shen2018, Gong2018, Zhou2019, Kawabata2019} and an unconventional bulk-boundary correspondence due to anomalous sensitivity to boundary conditions~\cite{Lee2016, Xiong2018, Kunst2018, Yao2018, Lee2019, Kunst2019, Yokomizo2019, Okuma2019} have been found with no counterparts in Hermitian systems. In many-body systems~\cite{Fendley1993A, Fendley1993B, Lee2014X, Lee2014L, Ashida2016, Ashida2017, Nakagawa2018, Hamazaki2019, Yoshida2019a, Guo2020}, non-Hermiticity can induce quasi-long-range ordered phases with power-law decaying correlations even without continuous symmetry in the Hamiltonian~\cite{Lee2014X}. %, and it also generates
Non-Hermiticity also leads to
unconventional renormalization-group flows that are forbidden in Hermitian systems~\cite{Fendley1993A, Fendley1993B, Ashida2017, Nakagawa2018}.
However, the crucial role of an energy gap in quantum phase transitions has yet to be %clarified 
fully understood
in non-Hermitian many-body systems.

In this Letter, we show that a continuous quantum phase transition can occur even without gap closing in non-Hermitian quantum many-body systems. 
%This implies that the transition is not necessarily associated with gap closing in non-Hermitian systems. %ver.3
In such a transition, the susceptibility, which is related to the spatial correlation and fluctuations of a local physical quantity, develops a singularity because of the nonorthogonality of eigenstates. This makes a sharp contrast with the Hermitian case, in which the singularity of the susceptibility originates from gap closing~\cite{You2007, Venuti2007, Gu2010}.
These facts imply that the %relation 
relationship
between the correlation length and the energy gap is fundamentally altered and the framework of continuous quantum phase transitions should be reconsidered %from scratch 
in non-Hermitian systems.
By way of illustration, we construct an exactly solvable non-Hermitian model by introducing non-Hermiticity to Kitaev's toric-code model~\cite{Kitaev2003b}.

%%%%%%%%%%
\paragraph{Breakdown of the Lieb-Robinson bound.\,---}

Under continuous deformation of a local and gapped Hermitian Hamiltonian $H(s)$, a change in the ground state $\ket{\psi_{0}(s)}$ can be described by a local unitary transformation $U(s)$, or a finite-time evolution generated by a local effective Hermitian Hamiltonian $\mathcal{D}(s)$~\cite{Hastings2005, Chen2010, Bravyi2010}. For a unique ground state, such a transformation is given by
\begin{align} \label{local-unitary}
\ket{\psi_{0}(s)} = U(s) \ket{\psi_{0}(0)},  
\end{align}
\begin{align}\label{passordered}
\quad U(s) := \mathcal{S'} \exp(\int_{0}^{s} i \mathcal{D}(s') \dd{s'}),
\end{align}
where ``$\mathcal{S'}\exp$'' denotes the $s'$-ordered exponential 
and $\mathcal{D}(s)$ is obtained from $H(s)$ as
\begin{align} \label{quasi-contin}
 i \mathcal{D}(s) = \int_{-\infty}^{\infty} F(t) \mathrm{e}^{i H(s) t} \qty(\dv{s} H(s)) \mathrm{e}^{- i H(s) t} \dd{t}.
\end{align}
The $s'$-ordered exponential in Eq.~\eqref{passordered} is defined by $\mathcal{S'} \exp(\int_{0}^{s} i \mathcal{D}(s') \dd{s'}) := \sum_{n=0}^{\infty} \frac{1}{n !} \mathcal{S'} \qty(\int_{0}^{s} i \mathcal{D}(s') \dd{s'})^{n}$,
where $\mathcal{S'}[\mathcal{D}(s'_{1}) \cdots \mathcal{D}(s'_{n})]$ is %given 
defined
by $\sum_{p \in S_{n}} \theta(s'_{p(1)} - s'_{p(2)}) \cdots \theta(s'_{p(n-1)} - s'_{p(n)}) \mathcal{D}(s'_{p(1)}) \cdots \mathcal{D}(s'_{p(n)})$ %with 
in terms of
the Heaviside unit-step function $\theta$.
In Eq.~\eqref{quasi-contin}, $F(t)$ is an odd function that decays faster than any negative power of $t$ for large $|t|$ and whose Fourier transform $\tilde{F}(\omega)$ is equal to $-1/\omega$ for $|\omega| > \Delta$ and infinitely differentiable. 
The presence of a finite gap $\Delta > 0$ guarantees that only $\omega$ with $|\omega| > \Delta$ matters, where $\tilde{F}(\omega)$ is smooth and $F(t)$ decays sufficiently fast~\cite{Chen2010, Bravyi2010}.

The locality of $\mathcal{D}(s)$ is guaranteed by the presence of a finite gap and the Lieb-Robinson bound~\cite{Lieb1972, Hastings2006, Nachtergaele2006}---the latter determines the speed limit $v_{\rm{L R}}$ with which an effective range of the support of a local operator $\dv{s}H(s)$ expands under a finite-time evolution generated by $H(s)$. 
One can restrict the action of the time-evolved operator to this effective range since the operator distance (i.e., the operator norm of the difference) between the original and restricted operators is negligibly small~\cite{Bravyi2006}. 
The integrand in Eq.~\eqref{quasi-contin} thus remains local for finite $t$, and only the integral over small $|t|$ is relevant because of the fast decay of $F(t)$, which is guaranteed by the presence of a finite gap as mentioned above. Owing to the locality of $\mathcal{D}(s)$, properties of $\ket{\psi_{0}(0)}$ with respect to a local operator $O$ are preserved under the local unitary transformation in Eq.~\eqref{local-unitary}. %Namely, the 
The
operator $ U^{\dagger}(s) O U(s)$ in the expectation value $\ev{O}{\psi_{0}(s)} = \ev{ U^{\dagger}(s) O U(s)}{\psi_{0}(0)}$ remains local because of the Lieb-Robinson bound~\cite{Chen2010, Bravyi2010}. Here an effective range of each local term in $\mathcal{D}(s)$ is estimated to be $\xi_{0} + v_{\rm LR} / \Delta$, where $\xi_{0}$ denotes the supremum of the interaction range (i.e., the diameter of the support of a local term in the Hamiltonian) of $H(s)$. Because of finite $v_{\rm LR}$, the locality of $\mathcal{D}(s)$ breaks down and a change in $\ket{\psi_{0}(s)}$ can be nonlocal only for $\Delta = 0$, which corresponds to a continuous phase transition.

In contrast, the Lieb-Robinson bound can, in general, break down in open-system dynamics conditioned on measurement outcomes~\cite{Ashida2018} %---the simplest example is 
such as
a non-Hermitian evolution corresponding to the null-jump process.
Let $H$ be a local non-Hermitian Hamiltonian $H = \sum_{Z} \qty( h_{Z}^{\rm H} + i h_{Z}^{\rm AH} )$, where $h_{Z}^{\rm H}$ and $i h_{Z}^{\rm AH}$ 
represent the Hermitian (H) and anti-Hermitian (AH) parts of the local term with support $Z$. We consider the time evolution of a local operator $O$ with support $X$: $O(t) = \exp(i H^{\dag} t) O \exp(-i H t)$.  
Then we have
\begin{align}
\eval{\dv{t} O(t)}_{t=0}  = \sum_{Z: Z \cap X \neq \emptyset} i \comm{h_{Z}^{\rm H}}{O} + \sum_{Z} \acomm{h_{Z}^{\rm AH}}{O}.
\end{align}
For the Hermitian parts $h_{Z}^{\rm H}$'s, commutators, which are taken with $O$, vanish for 
those $Z$'s that satisfy $Z \cap X =\emptyset$. For the anti-Hermitian parts, in contrast, anticommutators are taken with $O$; then, contributions from $h_{Z}^{\rm AH}$'s with $Z \cap X =\emptyset$ remain nonvanishing and affect the dynamics of $O$ directly, which indicates the breakdown of locality. 
To understand the physical origin, we consider the dissipative dynamics generated by a local Lindbladian $\mathcal{L}$~\cite{Lindblad1976}, which corresponds to the dynamics obtained after taking the ensemble average over all the possible measurement outcomes (i.e., quantum trajectories).
In the Heisenberg picture, such a dissipative dynamics is described by $ \eval{\dv{t} O(t)}_{t=0} = {\cal L}[O]$ with
\begin{align} \label{Lindblad}
\mathcal{L}[O] = \sum_{Z} \qty[  i \comm{h_Z}{O} + \sum_{j} \qty( {L^{j}_{Z}}^{\dagger} O L^{j}_{Z} - \frac{1}{2} \acomm{{L_{Z}^{j}}^{\dagger} L_{Z}^{j}}{O} ) ],
\end{align}
where $L_{Z}^{j}$'s are local jump operators with support $Z$. In the dynamics under continuous observation without quantum jumps,
the jump terms ${L^{j}_{Z}}^{\dagger} O L^{j}_{Z}$'s play no roles
and the effective non-Hermitian Hamiltonian is obtained %with 
as
$h_{Z}^{\rm H} = h_{Z}$ and $h_{Z}^{\rm AH} =  -\frac{1}{2} \sum_{j} {L_{Z}^{j}}^{\dagger} L_{Z}^{j}$. We note that %in Eq.~\eqref{Lindblad}, the sum 
the sum in Eq.~\eqref{Lindblad}
can be restricted to $Z$ with $Z \cap X \ne \emptyset$ since the quantum jump term ${L^{j}_{Z}}^{\dagger} O L^{j}_{Z}$ cancels $\frac{1}{2} \acomm{{L_{Z}^{j}}^{\dagger} L_{Z}^{j}}{O}$ for $Z \cap X = \emptyset$; this means the preservation of the locality of the dynamics, which results in the Lieb-Robinson bound in local Lindblad equations~\cite{Poulin2010, Nachtergaele2011, Barthel2012, Kliesch2014}. 
In contrast, when one considers the dynamics conditioned on measurement outcomes, %including 
such as
the non-Hermitian evolution, %as a specific example, 
the above cancellation does not occur in general and thus the Lieb-Robinson bound can be violated. 
This holds true even when a finite number of quantum jumps occur as long as a subensemble of quantum trajectories is of interest for continuous observation~\cite{Ashida2018}.

The breakdown of the Lieb-Robinson bound demonstrated above indicates that the correspondence between quantum phase transitions and gap closing can break down in non-Hermitian systems.
In fact, in the non-Hermitian case, $v_{\rm LR}$ has no general upper bound and thus the length scale $v_{\rm LR} / \Delta$ can diverge even without gap closing.

%%%%%%%%%%
\paragraph{Nonorthogonality-induced singularity.\,---}

\begin{figure}
 \centering
 \includegraphics[width=86mm]{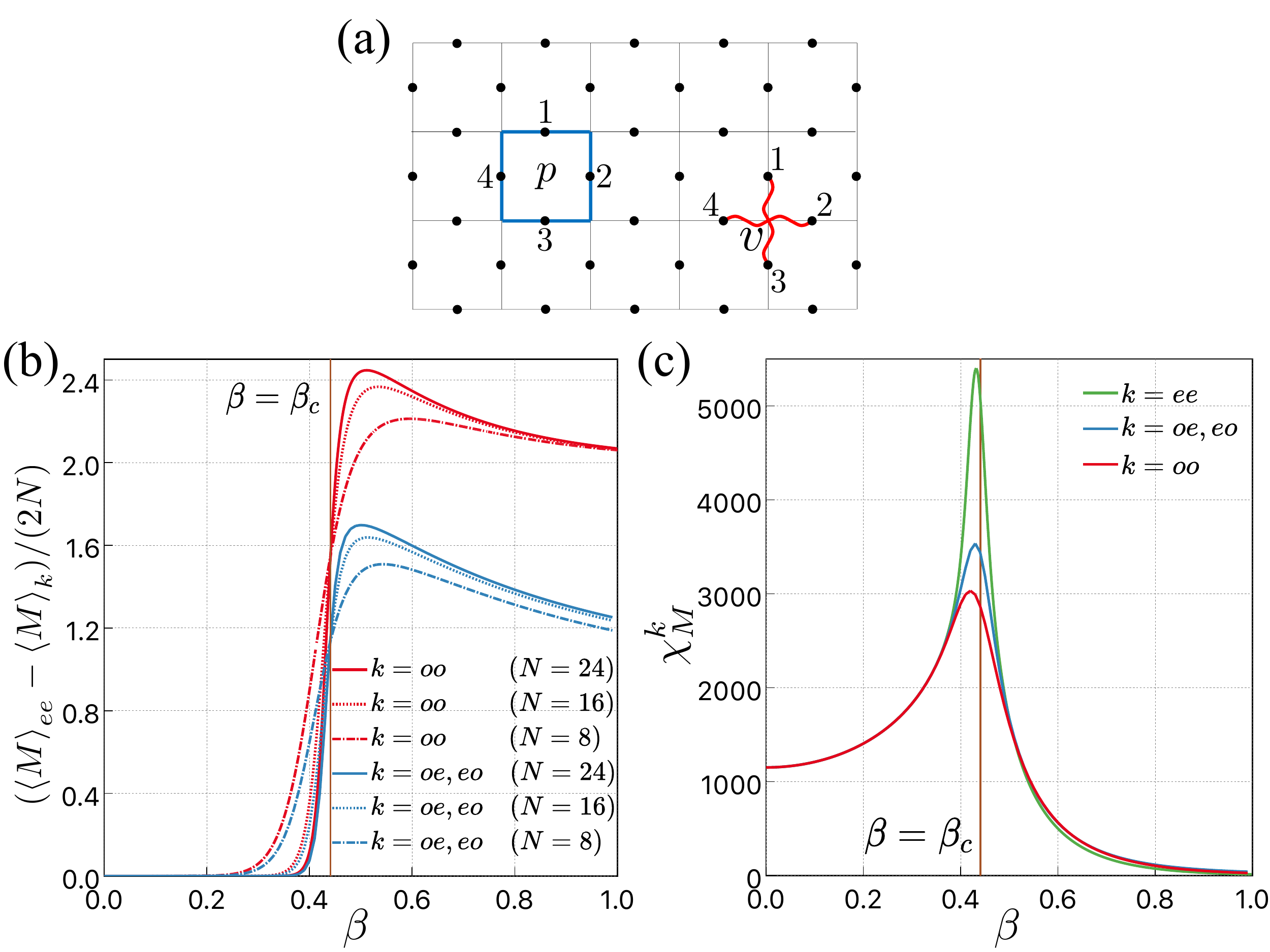}
 \caption{Non-Hermitian toric-code model.
(a) Spin-1/2 magnetic moments placed at the edges of a square lattice with $N \times N$ vertices. 
(b) Difference (scaled by $2N$) in the magnetization $\ev{M}_{k}(\beta)$ between different ground states for $N = 8, 16, 24$. Here $k \in \{ ee, eo, oe, oo \}$ labels different topological sectors, where the first (second) letter represents the parity of the number of noncontractible loops of down spins on the dual lattice winding around the torus in the $x$ ($y$) direction. 
%For $\beta<\beta_{c}$, the difference in $\ev{M}_{k}(\beta)$ tends to vanish with an increase in $N$, which suggests a topological phase.
%For $\beta > \beta_{c}$, the difference in $\ev{M}_{k}(\beta)$ becomes of the order of $N$, which suggests a trivial phase. 
For $\beta<\beta_{c}$ ($\beta > \beta_{c}$), the difference in $\ev{M}_{k}(\beta)$ tends to vanish (becomes of the order of $N$) with an increase in $N$, which suggests a topological (trivial) phase. %ver. 3
Moreover, the different curves cross at the transition point $\beta = \beta_{c}$.
(c) Magnetic susceptibility $\chi_{M }^{k} ( \beta )  = \frac { \mathrm { d } } { \mathrm { d } \beta } \langle  { M } \rangle _{k}( \beta )$ for $N=24$, which exhibits a singularity at $\beta = \beta_{c}$.} 
\label{figures}
\end{figure}

To gain further insight into the breakdown of the correspondence between quantum phase transitions and gap closing in non-Hermitian systems, we consider the fidelity susceptibility~\cite{You2007, Venuti2007, Gu2010}, which measures how rapidly the ground state changes under the variation of the system's parameter $\lambda$ and scales superextensively (i.e., grow more than extensively as a function of the system size) at a quantum phase transition reflecting long-range correlations.
We consider a non-Hermitian local Hamiltonian $H(\lambda) = H_{0} + \lambda V$, where $V := \sum_{i} V_{i}$ with $V_{i}$'s being local,
and let $\ket{\psi_{n}^{R}(\lambda)}$ and $\ket{\psi_{n}^{L}(\lambda)}$ denote the right and left eigenstates, respectively, with the (generally complex) eigenenergy $E_{n}(\lambda)$ and the normalization conditions $\braket{\psi_n^{R}(\lambda) | \psi_n^{R}(\lambda)} = 1$ and $\braket{\psi_m^{L}(\lambda) | \psi_n^{R}(\lambda)} = \delta_{m, n}$~\cite{Brody2014}. The right (left) eigenstates with different eigenenergies can be nonorthogonal, i.e., $\braket{\psi_{m}^{R (L)} (\lambda) | \psi_{n}^{R (L)} (\lambda) } \neq 0$ for $m \neq n$, owing to non-Hermiticity. We assume that the ground state $\ket{\psi_{0}^{R}(\lambda)}$ is unique with an excitation gap above it. Here, we define the ground state as the state with the lowest real part of the eigenenergy and the 
energy
gap as $\min_{n \ne 0} |E_{n}(\lambda) - E_{0}(\lambda) |$. 
We consider the fidelity $F(\lambda, \delta\lambda) := \qty| \braket{\psi_0^{R}(\lambda) | \psi_0^{R}(\lambda + \delta\lambda)} | $ for the right eigenstates~\footnote{In our normalization, other choices of fidelity such as $|\braket{\psi_{0}^{L}(\lambda) | \psi_{0}^{R}(\lambda + \delta\lambda)}|$ involve a term of the first order in $\delta\lambda$ in general, and the subsequent argument cannot 
straightforwardly %ver.2
be applied to them. %straightforwardly.
}. To the second order in $\delta \lambda$, we have~\footnote{See Supplemental Material for detailed discussions on the fidelity susceptibility, %and its upper bound, %change in ver.2-2
the calculation of the magnetization [Fig.~\ref{figures}(b)] and the magnetic susceptibility [Fig.~\ref{figures}(c)], the relationship between the magnetic and fidelity susceptibilities, an experimental situation of the non-Hermitian toric-code model, and topological quantum phase transitions in one-dimensional non-Hermitian systems. This Supplemental Material includes Refs.~\cite{Weinberg2015, Fisher1966, Kasteleyn1961, Kasteleyn1963, Fjaerestad2008, Samuel1980,  Verstraete2006B, Verstraete2006L, Schollwock2005, Fannes1992, Perez-Garcia2007}.}
\begin{align} \label{F2}
 F(\lambda, \delta\lambda) ^{2} = 1- \delta\lambda^{2} \braket{\partial_{\lambda} \psi_{0}^{R}(\lambda) | \partial_{\lambda} \psi_{0}^{R}(\lambda) }.
\end{align}
Hence the fidelity susceptibility is given by 
\begin{equation}
%\hspace{-2.63mm} 
 \chi _ { F} ( \lambda ) := \lim _ { \delta \lambda \rightarrow 0 } \frac { - 2 \ln F(\lambda, \delta\lambda)  } { \delta \lambda ^ { 2 } }  = \braket{\partial_{\lambda} \psi_{0}^{R}(\lambda) | \partial_{\lambda} \psi_{0}^{R}(\lambda) }.
 \end{equation}
Using the perturbation theory, we have
\begin{align} \label{NHchiF}
\chi_{F}(\lambda) = \sum_{m, n \ne 0} &\frac{\mel{\psi_{0}^{R}}{V^{\dag}}{\psi_{m}^{L}}}{\qty(E_{0} - E_{m})^{*}} \frac{\mel{\psi_{n}^{L}}{V}{\psi_{0}^{R}}}{
E_{0} - E_{n}} \\ \nonumber
 &\times \qty(\braket{\psi_{m}^{R} | \psi_{n}^{R}} - \braket{\psi_{m}^{R} | \psi_{0}^{R}} \braket{\psi_{0}^{R} | \psi_{n}^{R}}).
\end{align}

If the Hamiltonian is Hermitian, owing to the orthogonality of eigenstates, we have~\cite{You2007, Venuti2007, Gu2010}
\begin{align} 
\chi _ { F } ( \lambda ) = \sum_{n \ne 0} \frac{\left|\mel{\psi_{n}(\lambda)}{V}{\psi_{0}(\lambda)}\right|^{2}}{\left| E_{0} (\lambda) -E_{n} (\lambda)\right|^{2}}, \label{HermitianChiF} 
\end{align}
which can be rewritten as
\begin{align}
	\int_{0}^{\infty} \dd{\tau} \tau \sum_{i, j} 
	\qty[ \ev{V_{i}(\tau)  V_{j}(0) } - \ev{V_{i}(\tau)} \ev{V_{j}(0) } ], \label{Hint}
\end{align}
where $V_{i}(\tau) := e^{ H(\lambda) \tau } V_{i} e^{ - H(\lambda) \tau } $.
Here, the superscripts $L$ and $R$ are omitted since the left and right eigenstates are equivalent.
When the excitation gap closes, the denominator of the right-hand side of Eq.~\eqref{HermitianChiF} vanishes 
for some $n$ in the thermodynamic limit, which results in a superextensive scaling of $\chi_{F}$ and signals a quantum phase transition. If the gap is open, in contrast, correlations are short ranged and the summands in Eq.~\eqref{Hint} decay rapidly with distance, which is also guaranteed by the Lieb-Robinson bound~\cite{Hastings2004}; thus Eq.~\eqref{Hint} cannot grow superextensively~\cite{Venuti2007, Gu2010}.
This gives an alternative explanation for the correspondence between gap closing and a quantum phase transition in Hermitian systems. 

In non-Hermitian systems, however, the fidelity susceptibility can exhibit a superextensive scaling even without gap closing. This is because a large number of terms in the double sum in Eq.~\eqref{NHchiF} contribute to $\chi_{F}$ owing to the nonorthogonality of eigenstates in sharp contrast with the Hermitian case. In fact, Eq.~\eqref{NHchiF} can be rewritten as~\cite{Note2}
\begin{align}
\int_{-\infty} ^{0} \dd{\tau'} \int_{-\infty} ^{0} \dd{\tau} \sum_{i, j}
	\qty[	\ev{  V_{i}(\tau')^{\dag} V_{j}(\tau)  } - \ev{V_{i}(\tau')^{\dag} } \ev{V_{j}(\tau) } ].
\end{align}
This form looks similar to Eq.~\eqref{Hint} but can grow superextensively even if the energy gap is nonzero owing to the long-range correlations arising from the breakdown of the Lieb-Robinson bound.
% ver.3
%Here we emphasize that this superextensive scaling without gap closing has the physical origin different from that in Ref.~\cite{Cozzini2007}, in which the long-range coupling in the Hermitian models causes such scaling as well as the breakdown of the Lieb-Robinson bound.
%
Here we emphasize that this superextensive scaling without gap closing contrasts sharply with that found in a Hermitian model with a long-range coupling~\cite{Cozzini2007}; in the latter case, the breakdown of the Lieb-Robinson bound is caused by the long-range coupling. 
We also note that the breakdown of the Lieb-Robinson bound and the nonorthogonality of eigenstates are,
in general, %ver3
neither necessary nor sufficient to each other.

%%%%%%%%%%
\paragraph{Non-Hermitian toric-code model.\,---}

As an illustrative example, we consider the following non-Hermitian extension of Kitaev's toric-code model~\cite{Kitaev2003b}: 
\begin{equation}\label{NHmodel}
 H ( \beta ) = - \sum _ { v \in \{ \text {vertex} \} } A_{v}(\beta) - \sum _ { p \in \{ \text {plaquette} \} } B_{p},
\end{equation}
where $A_{v}(\beta) := \prod_{i=1}^{4} \sigma _ { v , i } ^ { \beta } $ and $B_{p} := \prod_{i=1}^{4} \sigma _ { p , i } ^ { z }$ are defined on four edges around a vertex $v$ and on a plaquette $p$ of a square lattice [Fig.~\ref{figures}(a)]. Here $\sigma^{x}_{i}, \sigma^{y}_{i}, \sigma^{z}_{i}$ are the Pauli matrices on the edge $i$, and the non-Hermitian operator $\sigma _ { i } ^ { \beta }$ is defined as 
\begin{align}
 \sigma _ { i } ^ { \beta } : = \cosh ( \beta ) \sigma _ { i } ^ { x } + i \sinh ( \beta ) \sigma _ { i } ^ { y } =\left( \begin{array}{cc}{0} & {\mathrm{e}^{\beta}} \\ {\mathrm{e}^{-\beta}} & {0}\end{array}\right), 
 \end{align}
where $\beta \ge 0$ parametrizes non-Hermiticity. This non-Hermitian operator physically represents an asymmetric spin flip. The original Hermitian model $H(0)$ is a prototypical solvable model that exhibits $\mathbb{Z}_{2}$ topological order~\cite{Kitaev2003b}. %As 
Since
all the terms appearing in the non-Hermitian Hamiltonian~\eqref{NHmodel} commute with one another, the exact solvability of the original model is maintained under the non-Hermitian extension. 

%%%%%%%%%%
%\paragraph{Basic Properties of the Ground States.\,---}

The original Hermitian model $ H(0) $ exhibits fourfold degenerate ground states below an excitation gap under the periodic boundary conditions (i.e., on a torus)~\cite{Kitaev2003b}. Importantly, the energy gap remains open in the presence of non-Hermiticity $\beta > 0$ as we explain in the following. Here $H(\beta)$ is related to the original Hermitian model $H(0)$ by a similarity transformation $ H ( \beta )  = S ( \beta ) H ( 0 ) S ( \beta ) ^ { - 1 }$, where $ S ( \beta ) : = \exp \left( \frac { \beta } { 2 } \sum _ { i } \sigma _ { i } ^ { z } \right)$. 
Thus, regardless of $\beta$, the energy spectrum remains unchanged in comparison with the Hermitian case, and there are fourfold degenerate ground states below the energy gap. 
The right (left) eigenstates with the eigenenergy $E_{n}$ are  $| \psi_{ n, k } ^ { R} ( \beta ) \rangle \propto S ( \beta ) | \psi _ { n, k }(0) \rangle$ [$\bra{\psi_{n, k}^{L}(\beta)} \propto \bra{\psi_{n, k}(0)} S^{-1}(\beta)$], where $k$ is the index labeling degenerate eigenstates and the superscripts $L, R$ are omitted for the Hermitian case ($\beta=0$). 
The fourfold ground states are superposition states of spin configurations $\{ \sigma_{i}^{z} \}$ in which down spins form closed loops on the dual lattice. For $\beta = 0$, such spin configurations are superposed with an equal weight within each topological sector characterized by the parities $(p_{x}, p_{y})$ of the numbers of noncontractible loops that wind around the torus in the $x$ and $y$ directions. As $\beta$ increases,
the weight of a configuration with a larger magnetization (i.e., a smaller total length of loops) becomes exponentially larger. %Eventually, for 
For $\beta \to \infty$, one of the ground states becomes fully polarized, and the topological feature is entirely lost.
In fact, a topological phase transition takes place at $\beta = \beta_{c} := ( 1 / 2 ) \ln ( \sqrt { 2 } + 1 )  \simeq 0.4407$~\cite{Castelnovo2008, Abasto2008}, as shown below.

%%%%%%%%%%
\paragraph{Topological phase transition.\,---}

A signature of topological order is given by topological entanglement entropy~\cite{Kitaev2006, Levin2006}, which is a subleading constant term $\gamma$ following the area-law term $\alpha L$ in the entanglement entropy $S$ for a subregion of the ground state: $S = \alpha L -\gamma + o(L^{0})$, where $L$ denotes the perimeter of the subregion and $\alpha$ is a constant.  In particular, the original Hermitian toric-code model has $\gamma = \ln2$~\cite{Hamma2005a, Hamma2005b}, which is a universal value for $\mathbb{Z}_{2}$ topological order.
Our non-Hermitian model possesses $\gamma = \ln2$ ($\gamma =0$) for $\beta < \beta_{c}$ ($\beta > \beta_{c}$), which indicates a topological (trivial) phase. To show this, we note that $H(\beta)$ shares the same ground states with the following Hermitian model: 
\begin{equation} \label{Castelnovo}
H =  -  \sum _ { p  } B _ { p } -  \sum _ { v  } A _ { v }  +  \sum _ { v } \exp( - \beta \sum _ { i \in v }  \sigma _ { i } ^ { z })
\end{equation} 
with $A_{v} := A_{v}(0)$. This Hermitian model was introduced in Ref.~\cite{Castelnovo2008} and $\gamma$ %was 
is
analytically obtained~\footnote{In Ref.~\cite{Castelnovo2008}, $\gamma$ is analytically obtained for one topological sector ($e e$). For the same topological sector, %the qualitative behavior of 
the magnetization and the magnetic susceptibility %is also derived. 
are discussed qualitatively.
These results are consistent with our quantitative %evaluation of them
results~[Fig.~\ref{figures} (b), (c)]} (see also Ref.~\cite{Stephan2009}).
Here, $\beta$ physically represents an external magnetic field for $|\beta| \ll 1$ in the Hermitian model \eqref{Castelnovo} while $\beta$ represents the degree of the asymmetric spin flips in our non-Hermitian model \eqref{NHmodel}. We note that gap closing at the transition point was numerically demonstrated in the former model~\cite{Isakov2011}, while the gap is constant regardless of $\beta$ in the latter one as similarity transformations do not alter the spectrum.

%%%%%%%%%%
%\paragraph{Local Observable.\,---}
Another important property of topological order is that the projection of any local operator onto the ground-state manifold is proportional to the identity: $\mel{\psi_{0, k'}}{{O}}{\psi_{0, k}} = c _ { {O} } \delta _ { k' , k }$~\cite{Ioffe2002, Bravyi2006, Furukawa2006, Furukawa2007, Nussinov2009, Schuch2010, Zeng2015}, which indicates that the degenerate ground states cannot be distinguished by any local observable. 
We examine this property for the total magnetization $M := \sum_{i} \sigma_{i}^{z}$.
Figure~\ref{figures}(b) shows the difference in the magnetization $\ev{M}_{k}(\beta)$ between different ground states, where $\ev{O}_{k}(\beta) := \bra{\psi_{0, k}^{R}(\beta)} O \ket{\psi_{0, k}^{R}(\beta)}$~\cite{Note2}.
%For $\beta<\beta_{c}$, the difference in $\ev{M}_{k}(\beta)$ tends to vanish with an increase in $N$, which indicates a topological phase. 
%For $\beta > \beta_{c}$, the difference is of the order of $N$, which indicates a trivial phase.
For $\beta<\beta_{c}$ ($\beta > \beta_{c}$), the difference in $\ev{M}_{k}(\beta)$ tends to vanish (becomes of the order of $N$) with an increase in $N$, which indicates a topological (trivial) phase. %ver. 3

The magnetic susceptibility $\chi_{M }^{k} ( \beta ) := \frac { \mathrm { d } } { \mathrm { d } \beta } \langle  { M } \rangle_{k}( \beta )$ exhibits a superextensive scaling at $\beta = \beta_{c}$ [Fig.~\ref{figures}(c)]~\cite{Note2}. For our model, this also indicates %the 
a
superextensive scaling of the fidelity susceptibility.
In fact, using the perturbation theory developed above with $V = \dv{\beta} H(\beta)$ and the fact that each excited state with a nonzero contribution to the sum in Eq.~\eqref{NHchiF} can be created by acting local operators as $\sigma_{i}^{z} \ket{\psi_{0, k}^{R}(\beta)}$, we can show
%\begin{align}
%\hspace{-3.035mm}
%\chi_{F}^{k}(\beta) &= \frac{1}{4} \sum_{i,j}  \qty(\ev{\sigma_{i}^{z} \sigma_{j}^{z}}_{k} - \ev{\sigma_{i}^{z}}_{k}\ev{\sigma_{j}^{z}}_{k})  = \frac{1}{4} \chi_{M}^{k}(\beta),
%\end{align}%
$\chi_{F}^{k}(\beta) = \frac{1}{4} \chi_{M}^{k}(\beta)$~\cite{Note2}, %ver.3
where $\chi_{F}^{k}(\beta)$ denotes the fidelity susceptibility for $\ket{\psi_{0, k}^{R}(\beta)}$~\footnote{The ground state is assumed to be unique in the general argument, whereas the ground states in this model are degenerate on a torus.
However, nondegenerate perturbation theory can still be applied within each topological sector where such degeneracy is removed.
The same result %was already 
is %ver.2
obtained for the Hermitian model \eqref{Castelnovo} by the direct calculation using the exact form of the ground state~\cite{Abasto2008}.}. 
Our model illustrates the superextensive scaling of the (fidelity) susceptibility due to the nonorthogonality of eigenstates as the energy gap remains nonvanishing for any $\beta$.
We note that such a transition cannot occur under similar transformations in one-dimensional non-Hermitian systems~\cite{Note2}.

%%%%%%%%%%
\paragraph{Experimental situation.\,---}

The non-Hermitian model in Eq.~\eqref{NHmodel} can be simulated experimentally with ultracold atoms. The dynamics by $H(\beta)$ can be decomposed as 
\begin{align}
 \mathrm{e}^{-i H(\beta) t}  = \qty( \prod_{p} \mathrm{e}^{i B_{p} t} ) S(\beta) \qty( \prod_{v} \mathrm{e}^{i A_{v} t} ) S^{-1}(\beta).
\end{align}
Schemes for simulating the unitary dynamics by $A_{v}$ or $B_{p}$ with ultracold atoms have been proposed in Refs.~\cite{Weimer2010, Weimer2011, Auger2017}, where the four-body interactions are simulated using the controlled-NOT gates~\cite{Muller2009} which can be implemented with Rydberg atoms~\cite{Gallagher1994, Tong2004, Saffman2016} and electromagnetically induced transparency~\cite{Boller1991, Fleischhauer2005, Weatherill2008}. 
Moreover, the nonunitary dynamics $S(\beta)$ and $S^{-1}(\beta)$ can be implemented by postselection of events without spontaneous decay of one of the spin components under continuous measurement~\cite{Lee2014L, Lee2014X}. All of these elements can be implemented, for example, with $^{87}$Rb atoms~\cite{Weimer2010, Lee2014L, Note2}. 

Physically, the ground state of $H(\beta)$ is a stationary state of the conditional dynamics, and can be prepared by starting with the ground state of the Hermitian counterpart at zero temperature and then adiabatically ramping up $\beta$~\cite{Ashida2017}. A signature of the proposed transition can be detected through a singularity in the magnetic susceptibility of the ground state.

 In summary, we have  demonstrated that continuous quantum phase transitions can occur without gap closing in non-Hermitian quantum many-body systems. In such a transition, the singularity of the fidelity susceptibility arises from nonorthogonality of eigenstates. 
Possible applications of our theory include adiabatic preparation of a topological phase from a trivial phase~\cite{Hamma2008} and an improved efficiency of quantum annealing~\cite{Finnila1994, Kadowaki1998, Johnson2011}.
The former can be realized by continuously changing the Hamiltonian without gap closing in a finite time via a process converse to that presented in this Letter---the presence of a gap helps to suppress nonadiabatic excitations.
For the latter, the annealing time, which is inversely proportional to the excitation gap by the adiabatic theorem, may be short even for a large non-Hermitian system. In Hermitian systems, by contrast, the energy gap is smaller and hence the operation time is longer for a larger system size.
In both examples, 
a %ver3
short operation time is also important in practice, since the probability of successful postselection under continuous measurement decays with time. 
It is worthwhile to explore concrete applications in these directions.

%%%%% Acknowledgement %%%%%
\smallskip
We are grateful to Zongping Gong, Masaya Nakagawa, Takashi Mori, Hosho Katsura, and Masaki Oshikawa for fruitful discussions. This work was supported by KAKENHI Grant No. JP18H01145 and a Grant-in-Aid for Scientific Research on Innovative Areas ``Topological Materials Science'' (KAKENHI Grant No. JP15H05855) from the Japan Society for the Promotion of Science (JSPS).
N.~M. was supported by the JSPS through Program for Leading Graduate Schools (MERIT). 
K.~K. was supported by JSPS KAKENHI Grant No. JP19J21927.
Y.~A. was supported by JSPS KAKENHI Grants No.~JP16J03613 and No.~JP19K23424.
S.~F. was supported by JSPS KAKENHI Grant No. JP18K03446 and Keio Gijuku Academic Development Funds.

\bibliography{NHTC8}

%\documentclass [prl, twocolumn, superscriptaddress]{revtex4-1} 
%\usepackage[dvipdfmx]{graphicx}
%\usepackage{graphicx}
%\usepackage{physics}
%\usepackage{bm, amsmath, amssymb, braket}
%\usepackage{times}
%\usepackage{multirow}
%\usepackage{ascmac}
%\usepackage{amsthm}
%\usepackage{float}
%\usepackage{color}
%\usepackage{comment}

%\begin{document}

\widetext
\pagebreak

\renewcommand{\theequation}{S\arabic{equation}}
\renewcommand{\thefigure}{S\arabic{figure}}
\renewcommand{\thetable}{S\arabic{table}}
\setcounter{equation}{0}
\setcounter{figure}{0}
\setcounter{table}{0}

\begin{center}
{\bf \large Supplemental Material for \\ \smallskip ``Continuous Phase Transition without Gap Closing\\ in Non-Hermitian Quantum Many-Body Systems"}
\end{center}

%%%%%%%%%%
\section{Fidelity susceptibility in non-Hermitian systems}

Here we derive %an expression of 
the fidelity susceptibility in non-Hermitian systems %using the perturbation theory. 
perturbatively.
%, and obtain an upper bound on it. 
% ver.2
To this end, 
we first expand $\ket{\psi_{0}^{R}(\lambda + \delta\lambda)}$ in powers of $\delta\lambda$ as
\begin{align}
\ket{\psi_{0}^{R}(\lambda + \delta\lambda)} = \ket{\psi_{0}^{R}(\lambda)} + \delta\lambda \ket{\partial_{\lambda} \psi_{0}^{R}(\lambda)} + \frac{\delta\lambda^{2}}{2} \ket{\partial^{2}_{\lambda} \psi_{0}^{R}(\lambda)} + \mathcal{O}(\delta\lambda^{3}).
\end{align}
Then the fidelity $F(\lambda, \delta\lambda) = \qty| \braket{\psi_0^{R}(\lambda) | \psi_0^{R}(\lambda + \delta\lambda)} | $ is expressed as
\begin{align} \label{F1} 
F(\lambda, \delta\lambda)^{2}
&= \qty|1 + \delta\lambda  \braket{\psi_{0}^{R}(\lambda) | \partial_{\lambda} \psi_{0}^{R}(\lambda)} + \frac{\delta\lambda^{2}}{2} \braket{\psi_{0}^{R}(\lambda) | \partial^{2}_{\lambda} \psi_{0}^{R}(\lambda)} + \mathcal{O}(\delta\lambda^{3})|^{2} \nonumber \\
&= 1 + \delta\lambda \, \partial_{\lambda} \qty( \braket{\psi_{0}^{R}(\lambda) | \psi_{0}^{R}(\lambda)}) \\ \nonumber
& \quad\quad + \delta\lambda^{2} \qty( \frac{1}{2} \braket{ \partial^{2}_{\lambda}\psi_{0}^{R}(\lambda) |  \psi_{0}^{R}(\lambda)} + \braket{ \partial_{\lambda} \psi_{0}^{R}(\lambda) |  \psi_{0}^{R}(\lambda)} \braket{\psi_{0}^{R}(\lambda) | \partial_{\lambda} \psi_{0}^{R}(\lambda)} +\frac{1}{2} \braket{\psi_{0}^{R}(\lambda) | \partial^{2}_{\lambda} \psi_{0}^{R}(\lambda)}) + \mathcal{O}(\delta\lambda^{3}). 
\end{align} 
%Using 
From
the normalization condition $\braket{\psi_{0}^{R}(\lambda) | \psi_{0}^{R}(\lambda)}=1$, we have
\begin{align} 
0= \partial_{\lambda} \qty( \braket{\psi_{0}^{R}(\lambda) | \psi_{0}^{R}(\lambda)}) 
= \braket{\partial_{\lambda}\psi_{0}^{R}(\lambda)| \psi_{0}^{R}(\lambda)} + \braket{\psi_{0}^{R}(\lambda)| \partial_{\lambda}\psi_{0}^{R}(\lambda)} = 2 \Re\qty[\braket{\psi_{0}^{R}(\lambda)| \partial_{\lambda}\psi_{0}^{R}(\lambda)} ]  \label{first}.
\end{align} 
Differentiating this equation with respect to $\lambda$, %again, 
we obtain
\begin{align} 
\braket{\partial^{2}_{\lambda} \psi_{0}^{R}(\lambda) | \psi_{0}^{R}(\lambda)} + 2\braket{\partial_{\lambda} \psi_{0}^{R}(\lambda) | \partial_{\lambda} \psi_{0}^{R}(\lambda)} + \braket{ \psi_{0}^{R}(\lambda) | \partial^{2}_{\lambda}\psi_{0}^{R}(\lambda)} =0 \label{second}.
\end{align}
%Substituting Eqs.~\eqref{first} and \eqref{second} into Eq.~\eqref{F1}, we have
Using Eqs.~\eqref{first} and \eqref{second}, we can rewrite Eq.~\eqref{F1} as 
\begin{align}
%F(\lambda, \delta\lambda)^{2} = 1 + \delta\lambda^{2} \qty(\Im\qty[\braket{\psi_{0}^{R}(\lambda)| \partial_{\lambda}\psi_{0}^{R}(\lambda)}]^{2} - \braket{\partial_{\lambda} \psi_{0}^{R}(\lambda) | \partial_{\lambda} \psi_{0}^{R}(\lambda)}) + \mathcal{O}(\delta\lambda^{3}).
F(\lambda, \delta\lambda)^{2} = 1 + \delta\lambda^{2} \qty(\qty[ \Im \braket{\psi_{0}^{R}(\lambda)| \partial_{\lambda}\psi_{0}^{R}(\lambda)}]^{2} - \braket{\partial_{\lambda} \psi_{0}^{R}(\lambda) | \partial_{\lambda} \psi_{0}^{R}(\lambda)}) + \mathcal{O}(\delta\lambda^{3}).
\end{align}
Here, %$\Im\qty[\braket{\psi_{0}^{R}(\lambda)| \partial_{\lambda}\psi_{0}^{R}(\lambda)}]$ 
$\Im \braket{\psi_{0}^{R}(\lambda)| \partial_{\lambda}\psi_{0}^{R}(\lambda)}$ 
is not uniquely determined from the normalization
and can be set to zero with an appropriate choice of the $U(1)$ phase degree of freedom~\cite{Weinberg2015}. With this convention (i.e., $\braket{\psi_{0}^{R}(\lambda)| \partial_{\lambda}\psi_{0}^{R}(\lambda)} =0$), we obtain
\begin{align}
F(\lambda, \delta\lambda)^{2} = 1- \delta\lambda^{2} \braket{\partial_{\lambda} \psi_{0}^{R}(\lambda) | \partial_{\lambda} \psi_{0}^{R}(\lambda) } + \mathcal{O}(\delta\lambda^{3}).
\end{align}
Hence the fidelity susceptibility is defined as
\begin{equation} \label{chiF}
 \chi _ { F} ( \lambda ) := \lim _ { \delta \lambda \rightarrow 0 } \frac { - 2 \ln F(\lambda, \delta\lambda)  } { \delta \lambda ^ { 2 } }  = \braket{\partial_{\lambda} \psi_{0}^{R}(\lambda) | \partial_{\lambda} \psi_{0}^{R}(\lambda) }.
 \end{equation}
 We can expand $\ket{\partial_{\lambda} \psi_{0}^{R}(\lambda)}$ using the right eigenstates of $H(\lambda)$ as
\begin{align} \label{expand}
\ket{\partial_{\lambda} \psi_{0}^{R}(\lambda)} = c_{0}^{(1)}(\lambda) \ket{\psi_{0}^{R}(\lambda)} + \sum_{n \ne 0} c_{n}^{(1)}(\lambda)\ket{\psi_{n}^{R}(\lambda)}.
\end{align}
For $n \ne 0$, using the perturbation theory, we have
\begin{align} \label{nne0}
c_{n}^{(1)}(\lambda)  =\frac{\mel{\psi_n^{L}(\lambda)}{V}{\psi_0^{R}(\lambda)}}{E_{0}(\lambda) - E_{n}(\lambda)}.
\end{align}
For $ n=0$, because of the %phase 
convention $\braket{\psi_{0}^{R}(\lambda)| \partial_{\lambda}\psi_{0}^{R}(\lambda)} =0$, we have 
\begin{align} \label{n0}
c_{0}^{(1)}(\lambda) = - \sum_{n \ne 0} c_{n}^{(1)}(\lambda) \braket{\psi_0^{R}(\lambda) | \psi_n^{R} (\lambda)}.
\end{align}
Substituting %Eqs.~\eqref{expand}, \eqref{nne0}, and \eqref{n0} 
Eq.~\eqref{expand} with the coefficients \eqref{nne0} and \eqref{n0}
into Eq.~\eqref{chiF}, we obtain
\begin{align} \label{chiF-final}
\chi_{F}(\lambda) = \sum_{m, n \ne 0} \frac{\mel{\psi_{0}^{R}}{V^{\dag}}{\psi_{m}^{L}}}{\qty(E_{0} - E_{m})^{*}} \frac{\mel{\psi_{n}^{L}}{V}{\psi_{0}^{R}}}{
E_{0} - E_{n}} 
\qty(\braket{\psi_{m}^{R} | \psi_{n}^{R}} - \braket{\psi_{m}^{R} | \psi_{0}^{R}} \braket{\psi_{0}^{R} | \psi_{n}^{R}}).
\end{align}
%In particular, if all the eigenenergies are real, we obtain an upper bound on $\chi_{F}(\lambda)$ as
%\begin{align}
%\chi_{F}(\lambda)  &= \sum_{m, n \ne 0} \frac{\mel{\psi_{0}^{R}}{V^{\dag}}{\psi_{m}^{L}}}{E_{0} - E_{m}} \frac{\mel{\psi_{n}^{L}}{V}{\psi_{0}^{R}}}{E_{0} - E_{n}} \qty(\braket{\psi_{m}^{R}|\psi_{n}^{R}} - \braket{\psi_{m}^{R}|\psi_{0}^{R}} \braket{\psi_{0}^{R}|\psi_{n}^{R}}) \nonumber \\ 
%& \le \frac{1}{\Delta^{2}} \sum_{m, n \ne 0} \qty(\mel{\psi_{0}^{R}}{V^{\dag}}{\psi_{m}^{L}} \braket{\psi_{m}^{R}|\psi_{n}^{R}} \mel{\psi_{n}^{L}}{V}{\psi_{0}^{R}} - \mel{\psi_{0}^{R}}{V^{\dag}}{\psi_{m}^{L}} \braket{\psi_{m}^{R}|\psi_{0}^{R}} \braket{\psi_{0}^{R}|\psi_{n}^{R}} \mel{\psi_{n}^{L}}{V}{\psi_{0}^{R}} ) \nonumber \\
%&= \frac{1}{\Delta^{2}} \qty[\bra{\psi_{0}^{R}} V^{\dag} \qty(I - \ketbra{\psi_{0}^{L}}{\psi_{0}^{R}}) \qty(I - \ketbra{\psi_{0}^{R}}{\psi_{0}^{L}}) V \ket{\psi_{0}^{R}} - \bra{\psi_{0}^{R}} V^{\dag} \qty(I - \ketbra{\psi_{0}^{L}}{\psi_{0}^{R}}) \ketbra{\psi_{0}^{R}}{\psi_{0}^{R}} \qty(I - \ketbra{\psi_{0}^{R}}{\psi_{0}^{L}}) V \ket{\psi_{0}^{R}}]  \nonumber \\
%&= \frac{1}{\Delta^{2}} \qty(\ev{V^{\dag}V}{\psi_{0}^{R}} -  \ev{V^{\dag}}{\psi_{0}^{R}} \ev{V}{\psi_{0}^{R}})  \nonumber \\
%&= \frac{1}{\Delta^{2}} \sum_{i, j} \qty(\ev{V_{i}^{\dag}V_{j}}{\psi_{0}^{R}} -  \ev{V_{i}^{\dag}}{\psi_{0}^{R}} \ev{V_{j}}{\psi_{0}^{R}}),
%\end{align}•%
%where $\Delta$ denotes the energy gap.
% Changes (5) in ver.2
In fact,  $\chi_{F}(\lambda)$ obtained here can be rewritten as
\begin{align}
\int_{-\infty} ^{0} \dd{\tau'} \int_{-\infty} ^{0} \dd{\tau} \sum_{i, j}
	\qty[	\ev{  V_{i}(\tau')^{\dag} V_{j}(\tau)  } - \ev{V_{i}(\tau')^{\dag} } \ev{V_{j}(\tau) } ],
\end{align}
where $V_{i}(\tau) := e^{ H(\lambda) \tau } V_{i} e^{ - H(\lambda) \tau } $. This can be shown as follows:
 \begin{align}
& \quad \int_{-\infty} ^{0} \dd{\tau'} \int_{-\infty} ^{0} \dd{\tau} \sum_{i, j}
	\qty[	\ev{  V_{i}(\tau')^{\dag} V_{j}(\tau)  } - \ev{V_{i}(\tau')^{\dag} } \ev{V_{j}(\tau) } ] \\
&= \int_{-\infty} ^{0} \dd{\tau'} \int_{-\infty} ^{0} \dd{\tau} 
 	\qty( \ev{ V(\tau')^{\dag} V(\tau) } -  \ev{ V(\tau')^{\dag} } \ev{ V(\tau) } ) \\
&= \int_{-\infty} ^{0} \dd{\tau'} \int_{-\infty} ^{0} \dd{\tau} \sum_{m, n} e^{ -\qty( E_{0} - E_{m} )^{*} \tau'  } e^{ -\qty( E_{0} - E_{n} ) \tau} \nonumber\\
	&\quad \times \qty[ \mel{\psi_{0}^{R}}{V^{\dag}}{\psi_{m}^{L}} \braket{\psi_{m}^{R}|\psi_{n}^{R}} \mel{\psi_{n}^{L}}{V}{\psi_{0}^{R}} - \mel{\psi_{0}^{R}}{V^{\dag}}{\psi_{m}^{L}} \braket{\psi_{m}^{R}|\psi_{0}^{R}}  \braket{\psi_{0}^{R}|\psi_{n}^{R}}  \mel{\psi_{n}^{L}}{V}{\psi_{0}^{R}}  ] \\
&= \sum_{m, n \ne 0} \frac{\mel{\psi_{0}^{R}}{V^{\dag}}{\psi_{m}^{L}}}{\qty( E_{0} - E_{m} )^{*}} \frac{\mel{\psi_{n}^{L}}{V}{\psi_{0}^{R}}}{\qty( E_{0} - E_{n} ) } \qty(\braket{\psi_{m}^{R}|\psi_{n}^{R}} -  \braket{\psi_{m}^{R}|\psi_{0}^{R}}  \braket{\psi_{0}^{R}|\psi_{n}^{R}}  )
= \chi_{F}(\lambda).
 \end{align}

\section{Calculation of the magnetization and the magnetic susceptibility}

Here we show how to calculate the magnetization in the ground state of the non-Hermitian toric-code model. 
The exact form of a ground state in the $\{ \sigma_{i}^{z} \}$ basis is
\begin{align}
\ket{\psi_{0, k}^{R}(\beta)} = \frac{1}{\sqrt{Z_{k}(\beta)}} \sum_{c \in C_{k}} \exp(\frac{\beta}{2} \sum_{i}\sigma_{i}^{z}(c)) \ket{c},
\end{align}
where $C_{k}$ is the set of spin configurations $c = \{ \sigma_{i}^{z} \}$ in which down spins form closed loops on the dual lattice with the parity $k \in \{ ee, eo, oe, oo \}$ of the number of noncontractible loops winding around the torus in the $x$ and $y$ directions, and $\sigma_{i}^{z}(c)$ is the eigenvalue of $\sigma_{i}^{z}$ for $\ket{c}$. Here $Z_{k}(\beta)$ is the normalization factor defined by
\begin{align}
Z_{k}(\beta) := \sum_{c \in C_{k}} \exp(\beta \sum_{i}\sigma_{i}^{z}(c)).
\end{align}
Thus the expectation value of the magnetization $M = \sum_{i} \sigma_{i}^{z}$ is calculated as
\begin{align}
\ev{M}_{k}(\beta) = \frac{\sum_{c \in C_{k}} \sum_{i}\sigma_{i}^{z}(c) \exp(\beta \sum_{i}\sigma_{i}^{z}(c))}{\sum_{c \in C_{k}} \exp(\beta \sum_{i}\sigma_{i}^{z}(c))} = \dv{\beta} \ln(Z_{k}(\beta)),
\end{align}
from which the magnetic susceptibility $\chi_{M}^{k}(\beta)$ is also obtained by further differentiation.

 \begin{figure}
 \centering
 \includegraphics[width=86mm]{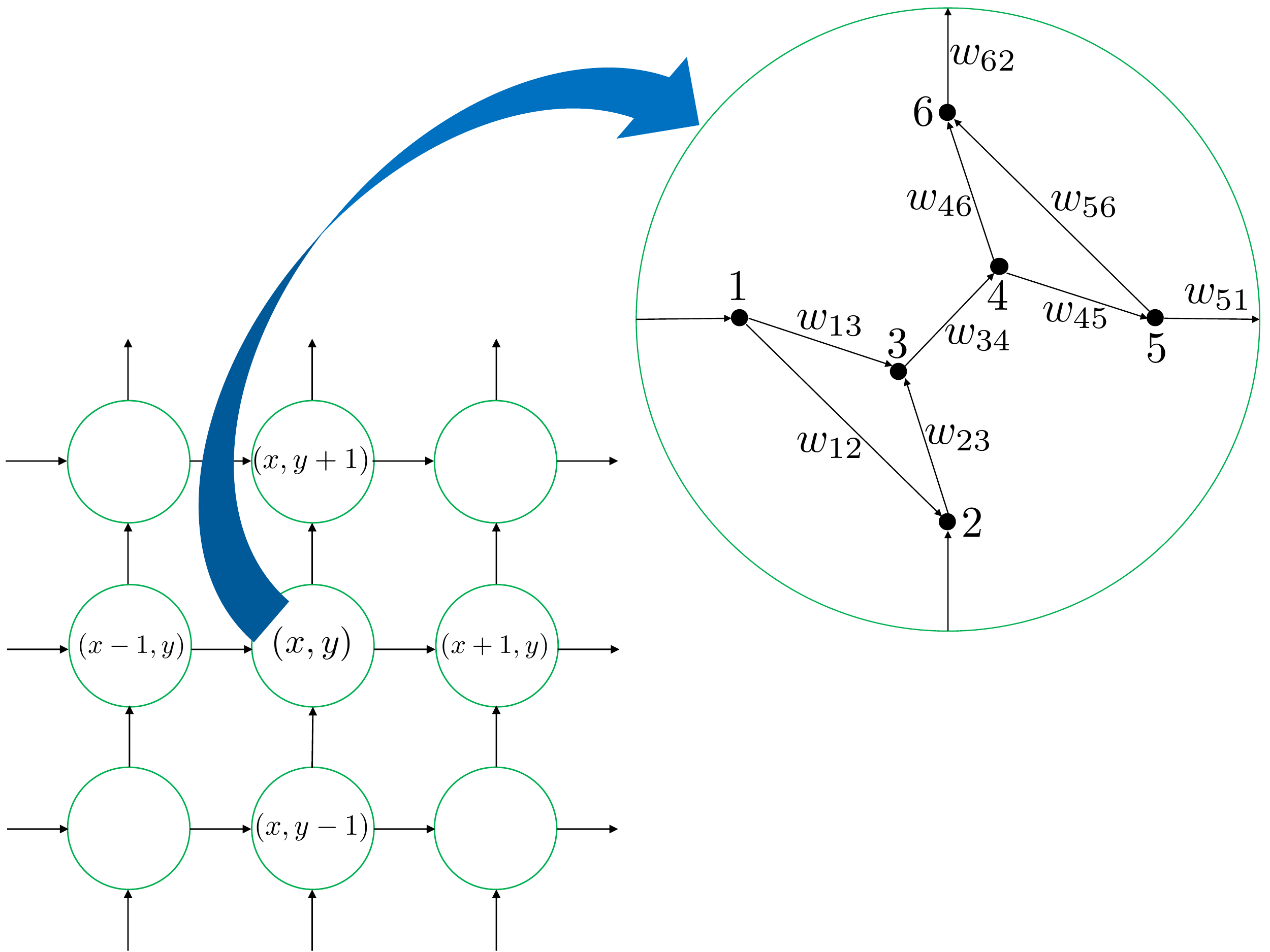}
 \caption{Star lattice. This lattice can be expressed as a square-lattice network of unit cells, each of which consists of six sites.
A site $i$ on this lattice can thus be expressed as a combination of an index $\alpha \in \{ 1, 2, \cdots 6\}$ within a unit cell and a coordinate $\vb{r} = (x, y)$ of a unit cell, where $x (y) \in \{1, 2, \cdots, N_{x} (N_{y}) \}$, and $N_{x}$ ($N_{y}$) denotes the number of unit cells in the $x$ ($y$) direction.  To calculate the magnetization in the non-Hermitian toric-code model, we make the weight for each edge depend only on the %degree 
degrees
of freedom within a unit cell: $w_{(x, y, \alpha),(x', y', \alpha')} (\beta) = w_{ \alpha, \alpha'} (\beta)$, where $w_{1 2} = w_{5 6} = \exp(-\beta), w_{1 3} = w_{2 3} = w_{4 5} = w_{4 6} = \exp(-\beta/2), w_{3 4} = 1, w_{5 1} = w_{6 2} = \exp(\beta)$.} 
\label{starlattice}
\end{figure}

\begin{figure}
 \centering
 \includegraphics[width=86mm]{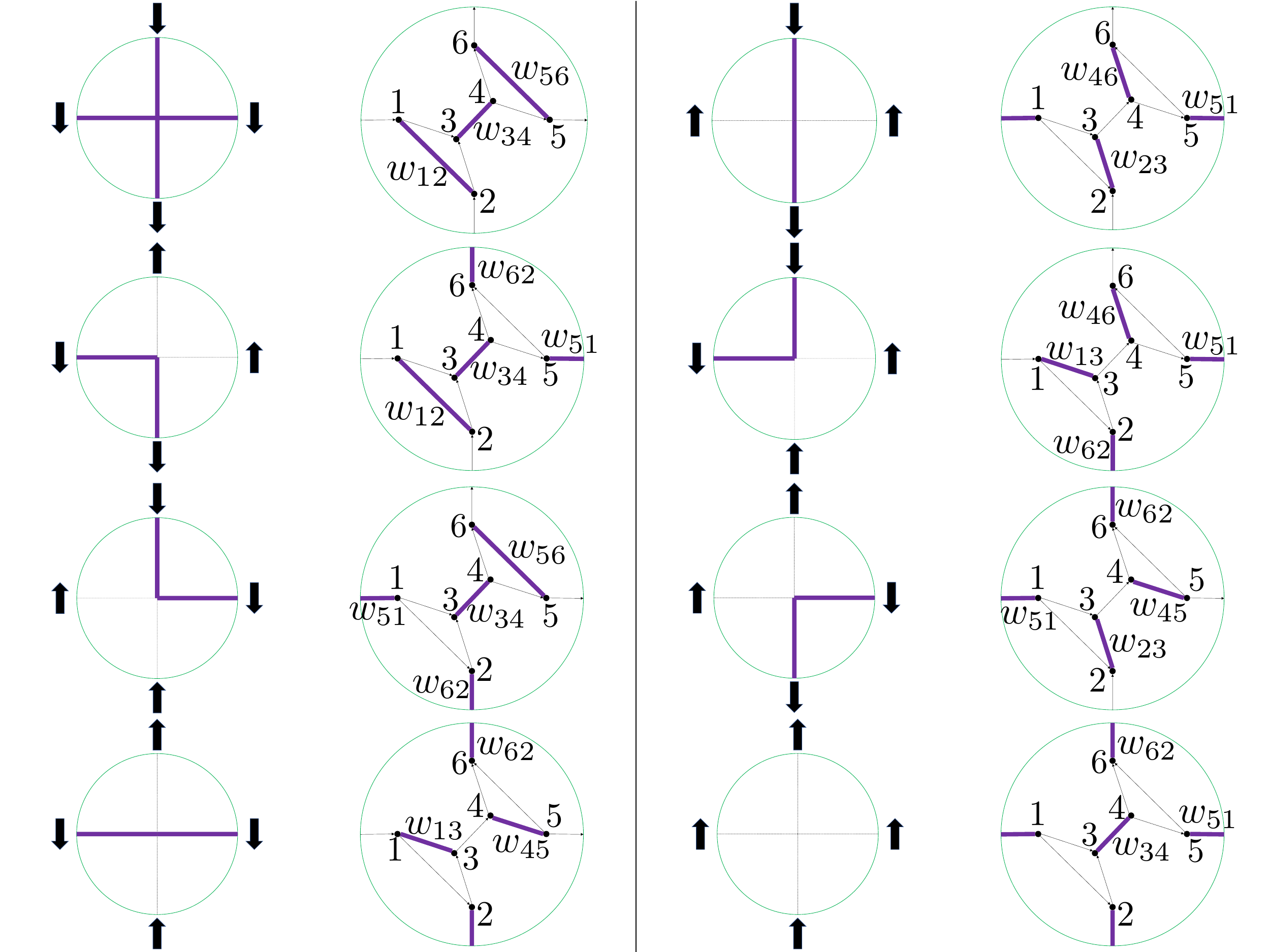}
 \caption{Correspondence between a closed-loop configuration on a square lattice and a dimer covering on a star lattice. %We note that 
 The square lattice on which closed loops of down spins are placed corresponds to the dual lattice in the original configuration of the system [Fig.~1(a) in the main text].} 
\label{correspondence}
\end{figure}

In the following, we calculate $Z_{k}(\beta)$, which can be regarded as a classical partition function for closed loop configurations. In fact, it is known that there is a one-to-one correspondence (Fig.~\ref{correspondence}) between a closed-loop configuration on a square lattice and a dimer covering on a star lattice~\cite{Fisher1966} (Fig.~\ref{starlattice}), for which the partition function can be calculated using the Pfaffian method~\cite{Kasteleyn1961, Kasteleyn1963, Fjaerestad2008}. 
A star lattice can be expressed as a square-lattice network of unit cells, each of which consists of six sites.
A site $i$ on this lattice can thus be expressed as a combination of an index $\alpha \in \{ 1, 2, \cdots 6\}$ within a unit cell and a coordinate $\vb{r} = (x, y)$ of a unit cell, where $x (y) \in \{1, 2, \cdots, N_{x} (N_{y}) \}$, and $N_{x}$ ($N_{y}$) denotes the number of unit cells in the $x$ ($y$) direction. For open boundary conditions, we define a Kasteleyn matrix $A(\beta)$, whose indices correspond to sites on a star lattice. The matrix elements of $A(\beta)$ are given by
\begin{align}
A_{i j}(\beta) =
	\begin{cases}
	\pm w_{i j}(\beta) & (\text{sites} \, i \,\text{and}\, j \, \text{are connected on a lattice}); \\
	0 & (\text{otherwise}),
	\end{cases}
\end{align}
where $w_{i j}(\beta)$ is the weight of a dimer placed on the edge between $i$ and $j$, and the sign is determined as follows: if the arrow is oriented from $i$ to $j$ in Fig.~\ref{starlattice}, the sign is positive, and otherwise negative. To calculate the magnetization in the non-Hermitian toric-code model, we make the weight for each edge depend only on the degree of freedom within a unit cell: $w_{(x, y, \alpha),(x', y', \alpha')} (\beta) = w_{ \alpha, \alpha'} (\beta)$, where $w_{1 2} = w_{5 6} = \exp(-\beta), w_{1 3} = w_{2 3} = w_{4 5} = w_{4 6} = \exp(-\beta/2), w_{3 4} = 1, w_{5 1} = w_{6 2} = \exp(\beta)$.
For periodic or anti-periodic boundary conditions, we define a Kasteleyn matrix $A^{(\nu_{x}, \nu_{y})}(\beta)$, where $\nu_{x}$ ($\nu_{y}$) corresponds to the boundary condition in the $x$ ($y$) direction as shown below. Compared with %the case of 
open boundary conditions, $A^{(\nu_{x}, \nu_{y})}(\beta)$ differs only in the presence of additional nonzero matrix elements involving sites along the boundaries: 
\begin{align}
^{\forall}y, \quad &A^{(\nu_{x}, \nu_{y})}_{(N_{x}, y, 5),(1, y, 1)}(\beta) = (-1)^{\nu_{x}} w_{(N_{x}, y, 5),(1, y, 1)}(\beta), \\
^{\forall}x, \quad &A^{(\nu_{x}, \nu_{y})}_{(x, N_{y}, 6),(x, 1, 2)}(\beta) = (-1)^{\nu_{y}} w_{(x, N_{y}, 6),(x, 1, 2)}(\beta),
\end{align}
with $A^{(\nu_{x}, \nu_{y})}_{i j}(\beta) = - A^{(\nu_{x}, \nu_{y})}_{j i}(\beta)$. Here the value $0$ ($1$) of $\nu_{x}$ and $\nu_{y}$ corresponds to the periodic (anti-periodic) boundary conditions in each direction. 
Then, $Z_{k}(\beta)$'s can be expressed as linear combinations of $\text{Pf} A^{(\nu_{x}, \nu_{y})}(\beta)$'s as~\cite{Kasteleyn1961}
\begin{align}
\mqty(Z_{e e}(\beta) \\ Z_{o e}(\beta) \\ Z_{e o}(\beta) \\ Z_{o o}(\beta) ) 
= \frac{1}{4} \mqty(1 & 1 & 1 & 1 \\ -1 & -1 & 1 & 1\\ -1 & 1 & -1 & 1\\ -1 & 1 & 1 & -1) \mqty(\text{Pf}A^{(0, 0)}(\beta) \\ \text{Pf}A^{(0, 1)} (\beta)\\ \text{Pf}A^{(1, 0)} (\beta)\\ \text{Pf}A^{(1, 1)}(\beta) ).
\end{align}

In the following, we calculate $\text{Pf}A^{(\nu_{x}, \nu_{y})}(\beta)$'s. We introduce a Grassmann variable $\psi_{i}$ for each site $i$. Then we define an action as
\begin{align}
S^{(\nu_{x}, \nu_{y})}(\beta) := \frac{1}{2} \sum_{i, j} \psi_{i} \, A^{(\nu_{x}, \nu_{y})}_{i j}(\beta) \, \psi_{j}.
\end{align}
With this action, $\text{Pf}A^{(\nu_{x}, \nu_{y})}(\beta)$ is expressed as the corresponding partition function~\cite{Samuel1980}:
\begin{align}
\int \mathcal{D}\psi \exp(S^{(\nu_{x}, \nu_{y})}(\beta)) = \text{Pf}A^{(\nu_{x}, \nu_{y})}(\beta).
\end{align}
We introduce new Grassmann variables $\tilde{\psi}_{\vb{q}, \alpha}$'s through the Fourier transform as
\begin{align}
\psi_{\vb{r}, \alpha} = \frac{1}{\sqrt{N_{x} N_{y}}} \sum_{\vb{q}} \mathrm{e}^{i \vb{q} \vdot \vb{r}} \tilde{\psi}_{\vb{q}, \alpha},
\end{align}
where $q_{x (y)} = 2\pi \qty(n_{x (y)} + \nu_{x (y)}/2) / N_{x (y)}$ and $n_{x (y)} \in \{0, 1, \cdots , N_{x (y)}-1 \}$. Then the action is rewritten as
\begin{align}
S^{(\nu_{x}, \nu_{y})}(\beta) = \frac{1}{2} \sum_{\vb{q}} \sum_{\alpha, \alpha'} \tilde{\psi}_{-\vb{q}, \alpha} \, \tilde{A}^{(\nu_{x}, \nu_{y})}_{\alpha \alpha'} (\vb{q}, \beta) \, \tilde{\psi}_{\vb{q}, \alpha'},
\end{align}
where $\tilde{A}^{(\nu_{x}, \nu_{y})} (\vb{q}, \beta) $ is a $6 \times 6$ matrix
\begin{align}
\tilde{A}^{(\nu_{x}, \nu_{y})} (\vb{q}, \beta)
 = \mqty( 0 & w_{1 2}(\beta) & w_{1 3}(\beta) & 0 & - w_{5 1}(\beta)\mathrm{e}^{-i q_{x}} & 0 \\
 		-w_{1 2}(\beta) & 0 & w_{2 3} (\beta)& 0 & 0 & - w_{6 2}(\beta)\mathrm{e}^{-i q_{y}} \\
		-w_{1 3}(\beta) & - w_{2 3}(\beta) & 0 & w_{3 4}(\beta) & 0 & 0 \\
		0 & 0 & -w_{3 4}(\beta) & 0 & w_{4 5}(\beta) & w_{4 6}(\beta) \\
		w_{5 1}(\beta)\mathrm{e}^{i q_{x}} & 0 & 0 &- w_{4 5}(\beta) & 0 & w_{5 6}(\beta) \\
		0 & w_{6 2}(\beta)\mathrm{e}^{i q_{y}} & 0 &- w_{4 6}(\beta) & -w_{5 6}(\beta) & 0
		).
\end{align}
Then we obtain $\text{Pf} A^{(\nu_{x}, \nu_{y})}(\beta)$ using the following relation:
\begin{align}
\text{Pf} A^{(\nu_{x}, \nu_{y})}(\beta)^{2}
= \det A^{(\nu_{x}, \nu_{y})}(\beta) 
= \prod_{\vb{q}} \det \tilde{A}^{(\nu_{x}, \nu_{y})}(\vb{q}, \beta).
\end{align}
This expression is used in our numerical calculation.

\section{Relationship between the fidelity and magnetic susceptibilities}

In the non-Hermitian toric-code model in Eq.~(12) in the main text, the fidelity susceptibility in a ground state is proportional to the magnetic susceptibility. To show this, we use the perturbation theory 
developed in the main text with $V = \dv{\beta} H(\beta) = \frac{1}{2} \qty(\sum_{i}\sigma_{i}^{z} H(\beta) - H(\beta) \sum_{i} \sigma_{i}^{z})$. 
For each ground state $\ket{\psi_{0, k}^{R}(\beta)}$ and $n \ne 0$, we have
\begin{align} \label{mel}
\frac{\mel{\psi_{n}^{L}(\beta)}{V}{\psi_{0, k}^{R}(\beta)}}{E_{0}(\beta) - E_{n}(\beta)} 
= \frac{1}{2} \sum_{i} \mel{\psi_{n}^{L}(\beta)}{\sigma_{i}^{z}}{\psi_{0, k}^{R}(\beta)}
=
	\begin{cases}
	\frac{1}{2} & \qty(\text{for} \, n \, \text{such that}\, ^{\exists}i, \ket{\psi_{n}(0)} = \sigma_{i}^{z} \ket{\psi_{0, k}(0)}); \\
	0 & \qty(\text{otherwise}).
	\end{cases}
\end{align}
Substituting Eq.~\eqref{mel} into Eq.~\eqref{chiF-final}, we have
\begin{align}
\chi_{F}^{k}(\beta) 
&= \frac{1}{4} \sum_{i,j}  \qty[ \qty(\bra{\psi_{0, k}^{R}(\beta)}\sigma_{i}^{z}) \qty(\sigma_{j}^{z} \ket{\psi_{0, k}^{R}(\beta)}) - \qty(\bra{\psi_{0,k}^{R}(\beta)}\sigma_{i}^{z})\ket{\psi_{0, k}^{R}(\beta)} \bra{\psi_{0, k}^{R}(\beta)}\qty(\sigma_{j}^{z}\ket{\psi_{0, k}^{R}(\beta)}) ] \\
&= \frac{1}{4} \sum_{i,j}  \qty(\ev{\sigma_{i}^{z} \sigma_{j}^{z}}_{k}(\beta) - \ev{\sigma_{i}^{z}}_{k}(\beta) \ev{\sigma_{j}^{z}}_{k}(\beta))  
= \frac{1}{4} \chi_{M}^{k}(\beta),
\end{align}%
where $\chi_{F}^{k}(\beta)$ and $\chi_{M}^{k}(\beta)$ denote the fidelity and magnetic susceptibilities, respectively, for $\ket{\psi_{0, k}^{R}(\beta)}$.

\section{Experimental situation}

The non-Hermitian toric-code model in Eq.~(12) can be simulated experimentally, for example, with $^{87}$Rb cold atoms. In this case, the spin states $\ket{\uparrow}$ and $\ket{\downarrow}$ correspond to $5S_{1/2}, F=1, m = \pm 1$ ground states~\cite{Lee2014L}.
From $ H ( \beta )  = S ( \beta ) H ( 0 ) S ( \beta ) ^ { - 1 }$, 
the dynamics governed by $H(\beta)$ can be decomposed as 
\begin{align}
 \mathrm{e}^{-i H(\beta) t}  
&= S(\beta) \mathrm{e}^{-i H(0) t} S^{-1}(\beta) \\
&= \qty( \prod_{p} \mathrm{e}^{i B_{p} t} ) S(\beta) \qty( \prod_{v} \mathrm{e}^{i A_{v} t} ) S^{-1}(\beta).
\end{align}
Some schemes for simulating the unitary dynamics by $A_{v}$ or $B_{p}$ with cold atoms have been proposed in Refs.~\cite{Weimer2010, Weimer2011, Auger2017}, where the four-body interaction in $A_{v}$ or $B_{p}$ is simulated using a controlled-NOT gate~\cite{Muller2009} implemented with Rydberg atoms~\cite{Gallagher1994, Tong2004, Saffman2016} and electromagnetically induced transparency~\cite{Boller1991, Fleischhauer2005, Weatherill2008}. In particular, for $^{87}$Rb atoms, we can use the $59s$ and $53s$ states as the Rydberg states for the control and target qubits,
respectively, in the above implementation~\cite{Weimer2010}.

Moreover, the nonunitary dynamics $S(\beta)$ and $S^{-1}(\beta)$ can be implemented by postselecting the event without a spontaneous decay of one of the spin components under continuous measurement~\cite{Lee2014L, Lee2014X}. This implementation requires %the use of 
an auxiliary state $\ket{a}$. In the case of $^{87}$Rb atoms,  we can use the $5S_{1/2}, F=2$ state as $\ket{a}$~\cite{Lee2014L}. The $\ket{\uparrow}$ state is optically pumped into the $\ket{a}$ state, and the population in the $\ket{a}$ state is continuously measured. In the absence of the population in the $\ket{a}$ state, the %conditioned 
conditional
dynamics is described as $\exp(- i H_{\rm eff} t)$, where $H_{\rm eff} = -(i/2) \sum_{j} L^{\dag}_{j} L_{j}$, and $L_{j} = \ketbra{a}{\uparrow}_{j}$ corresponds to the decay for the atom $j$. Then we have
\begin{align}
\exp(-i H_{\rm eff} t) = \exp(-\frac{t}{4} \sum_{j} \qty( 1 + \sigma_{j}^{z})).
\end{align}
Taking $t = 2\beta$, we simulate $S^{-1}(\beta)$ by $\exp(-i H_{\rm eff} (2\beta) )$. Meanwhile, using the relation $S(\beta)=\qty(\prod_{j} \sigma_{j}^{x}) S^{-1}(\beta)\qty(\prod_{j} \sigma_{j}^{x})$, we can also simulate $S(\beta)$ by combining $S^{-1}(\beta)$ with spin flip operations. 

In addition to a simulation of the dynamics, we can also experimentally realize the ground state of the non-Hermitian toric-code model $H(\beta)$
in the following manner. %ver.2
%To realize the ground state $\ket{\psi_{0, k}^{R} (\beta)}$ of the non-Hermitian toric-code model, we use the relation $\ket{\psi_{0, k}^{R} (\beta)} \propto S(\beta) \ket{\psi_{0, k} (0)} $. ver.2
We first prepare the ground state of the Hermitian model $\ket{\psi_{0, k} (0)} $ using the method proposed in Refs.~\cite{Weimer2010, Weimer2011}, and then %apply $S(\beta)$ as described above.  ver.2
adiabatically ramping up the strength of non-Hermiticity (see, for example, Supplementary Note 3 in Ref.~\cite{Ashida2017}), which corresponds to $\beta$ in our non-Hermitian toric-code model.

\section{Topological quantum phase transitions in one-dimensional non-Hermitian systems}

The topological quantum phase transition occurs at a finite parameter $\beta$ of non-Hermiticity 
under an invertible on-site nonunitary transformation % (added in ver.2)
in the non-Hermitian toric-code model. By contrast, such a transition cannot occur in one dimension as we explain in the following. For a one-dimensional local and gapped Hermitian Hamiltonian, the ground state can be well approximated by a matrix product state (MPS)~\cite{Verstraete2006B}, whereas the ground state of the Hermitian toric-code model can be expressed as a projected entangled pair state (PEPS)~\cite{Verstraete2006L}. In both cases, starting from the ground state of the Hermitian case, the system remains to be in an MPS or a PEPS under an 
invertible % (added in ver.2)
on-site nonunitary transformation $S$ 
because we can redefine tensors including the effect of the nonunitary transformation: $A_{i j}^{\alpha} \to \tilde{A}_{i j}^{\alpha} := S_{\alpha \beta} A_{i j}^{\beta}$, where $i$ and $j$ denote the virtual indices, and $\alpha$ and $\beta$ denote the physical indices for an MPS. 
An essential difference arises in the transfer matrix calculation of correlation functions of local operators. 
% (The following is modified in ver.2)
A PEPS can exhibit quasi-long-range correlations and describe a critical state even with a finite bond dimension $D$~\cite{Verstraete2006L}. 
In contrast, the correlation length is finite in an MPS 
and no phase transition accompanied by quasi-long-range correlations occurs if the maximum modulus of eigenvalues of a transfer matrix remains nondegenerate during the transformation~\cite{Schollwock2005}.
In fact, this is the case as shown in the following.
Since the original state satisfies the condition of nondegeneracy, a set of the corresponding tensors $\{A^{\alpha} \}$ satisfies injectivity (i.e., $\{A^{\alpha_{1}} \cdots A^{\alpha_{n}} \}$ spans the entire space of $D\times D$ matrices for some $n$)~\cite{Fannes1992, Perez-Garcia2007}.
%Such 
This
completeness of $\{A^{\alpha_{1}} \cdots A^{\alpha_{n}} \}$ is preserved under an invertible transformation $S$, and thus the collection of the transformed tensors remains injective, which guarantees the above-mentioned nondegeneracy. 

%\bibliography{NHTC4}

%\end{document}

\end{document}